\documentclass[letterpaper, 10 pt, conference]{ieeeconf}  
\IEEEoverridecommandlockouts                              
\usepackage{amsmath,amssymb,latexsym,mathrsfs,amssymb} 
\usepackage{array}
\usepackage{hyperref}
\usepackage[dvipsnames]{xcolor}
\usepackage{xfrac}
\usepackage{algorithm}
\usepackage{algorithmic}
\let\labelindent\relax
\usepackage{enumitem}
\usepackage{comment}
\usepackage[skins,breakable]{tcolorbox}

\title{\LARGE \bf
Competition and Cooperation of LLM Agents in Games
}

\author{Jiayi Yao, Cong Chen and Baosen Zhang
\thanks{J. Yao and B. Zhang are with the Department of Electrical and Computer Engineering, University of Washington, Seattle.  \{jiayiyao,zhangbao\}@uw.edu}
\thanks{C. Chen is with the Thayer School of Engineering at Dartmouth College. Cong.Chen@dartmouth.edu}
\thanks{This work has been submitted to the IEEE for possible publication. Copyright may be transferred without notice, after which this version may no longer be accessible.}
}

\begin{document}

\maketitle
\thispagestyle{empty}
\pagestyle{empty}

\begin{abstract}
Large language model (LLM) agents are increasingly deployed in competitive multi-agent settings, raising fundamental questions about whether they converge to equilibria and how their strategic behavior can be characterized. In this paper, we study LLM agent interactions in two standard games: a network resource allocation game and a Cournot competition game. Rather than converging to Nash equilibria, we find that LLM agents tend to cooperate when given multi-round prompts and non-zero-sum context. Chain-of-thought analysis reveals that fairness reasoning is central to this behavior. We propose an analytical framework that captures the dynamics of LLM agent reasoning across rounds and explains these experimental findings.
\end{abstract}

\section{Introduction}
Recent advances in large language models (LLMs) have given rise to a new class of decision makers in competitive settings~\cite{kang2026reasonably}. LLM-based agents are now being tasked to interact in dynamic environments ranging from economics to robotics to energy systems~\cite{huang2024understanding,zhang2025grid,jia2025enhancing,kannan2024smart}. In these settings, agents need to make strategic decisions while adapting to the behavior of others. Two natural questions arise: would the agents converge to an equilibrium, and if so, can we characterize this equilibrium? 

The question of the strategic behavior of agents has recently attracted considerable attention, although the results are somewhat scattershot. In some markets, studies have shown that the interaction of agents can lead to higher prices than what a competitive equilibrium would suggest~\cite{fish2024algorithmic}. However, in other studies, agents using popular standard models exhibit wildly inconsistent behaviors that are more similar to random actions than to strategic decision making~\cite{goodyear2025effect,hua2024game,buscemi2025fairgame}. The underlying issue does not seem to be the ability of agents to solve optimization problems. If the prompt is explicit, for example, ``play the best response with respect to this utility function'', most LLM models are able to follow this instruction. But this largely takes out strategic thinking out of the agents' hands and somewhat removes the appeal of using LLM agents. 

In practice, human users might simply instruct the agents to optimize their utilities over some time horizon without being very explicit about how to achieve this. In this paper, we are interested in a setting in which the agents are neither fully cooperative nor fully competitive (zero-sum games). In particular, we study two commonly used games: a network resource allocation game and a Cournot game.

The network resource allocation game was first introduced in~\cite{kelly1997charging}, where a network manager tries to allocate network capacity efficiently among users, each having a
utility function depending on their allocated rate.  If users are price-takers, then a proportional rate allocation mechanism achieves optimal efficiency~\cite{hajek2002greedy,stoica1996proportional}. In contrast, strategic price anticipatory users can worsen efficiency~\cite{johari2004efficiency}. We are interested in where LLM-agents would fall in this spectrum. 

Cournot games are another type of competition that has been studied extensively in the context of energy systems~\cite{kirschen2018fundamentals,zhang2015competition}, transportation~\cite{bimpikis2019cournot} and other settings. In this game, agents control their production, which in turn influences the price, and each receives a payment (or pays if they are modeled as buyers) based on the product of quantity and price. This type of competition is beginning to take shape in electricity markets, where generators, batteries, and large buildings are beginning to be managed by LLM agents~\cite{li2026behavioral,lu2024large,chen2025behavioral}. The efficiency of Cournot games has also been studied extensively~\cite{shi2020multi,johari2005efficiency}. 

In this paper, we show that in both of these games, LLM agents tend to cooperate with each other if given the opportunity. By this, we mean that if agents are prompted to maximize their net utility of multiple rounds (see Sec.\ref{sec:prompt}, and if the agents are given the context that the games are not zero-sum, they will try to maximize the total net utility and do not play towards Nash equilibria. Through a chain-of-thought analysis, we show that fairness plays a central role in their reasoning. We provide an analytical framework that explains the experimental behaviors and show how it models the dynamics of LLM agents reasoning process. 

As with many existing works on LLM agents, our results depend on the prompt and context given to the agents, as well as the exact LLM models used. But we believe that it provides a way to analytically study the behavior of LLM agents and can be applied to many engineering settings. This paper is organized as follows: Sec.~\ref{sec:model} describes the models used in this paper, Sec.~\ref{sec:prompt} provides the prompts given to the LLM agents, Sec.~\ref{sec:payoff} shows the payoff functions used by the agents, Sec.~\ref{sec:dynamics} shows how these functions evolve over multiple rounds and Sec.~\ref{sec:simulation} shows the simulation results.

\section{Setup} \label{sec:model}

In this section, we define the environments used to evaluate the behavior of Large Language Model (LLM) agents. We deploy two multi-agent environments: the Kelly mechanism for resource allocation~\cite{kelly1997charging,johari2004efficiency} and a Cournot game~\cite{friedman1982oligopoly}. In both environments, $N$ LLM agents interact simultaneously over $T$ finite time steps, aiming to maximize their cumulative net utility. At each round $t$, agents independently compute their actions based on the historical aggregate market state provided by the environment, without explicit knowledge of the number of opponents or their utilities. 

\subsection{Resource Allocation Game (Kelly Mechanism)}
Following the proportional resource allocation game described in~\cite{kelly1997charging}, we implement the game with the setting where a single divisible resource of fixed capacity $C > 0$ is shared among $N$ agents.
Each agent $i$  submits a non-negative bid $x_i \ge 0$ to the market operator, and the total sum of bids is $X = \sum_{j=1}^N x_j$. We sometimes use $X_{-i}=\sum_{j \neq i} x_j$ to denote the total sum of other bids. The mechanism allocates the rate $d_i$ to agent $i$ proportional to their bid: $d_i = \frac{x_i}{X} C$. The price per unit of the resource is $p = \frac{X}{C}$. Assuming each agent $i$ receives a utility $U_i(d_i)$ from their allocated rate, the net payoff for agent $i$ is:
\begin{equation}
    \pi_i = U_i\left( \frac{x_i}{x_i + \sum_{j \neq i} x_j} C \right) - x_i \label{eq:kelly_payoff}
\end{equation}
 As shown in~\cite{johari2004efficiency}, selfish price-anticipating behavior in this game can lead to loss in efficiency (bounded by a factor of $3/4$). To achieve the social optimal outcome, agents must coordinate in some way. We use this game to test whether the LLM agents would (implicitly) attempt to coordinate or play selfishly.

\subsection{The Cournot Game}
The Cournot game is a fundamental model of quantity competition where $N$ firms simultaneously decide their production levels. Following the framework in~\cite{johari2005efficiency}, we specifically consider the linear Cournot model. The market price is determined by an affine function of the total quantity, $p(x) = \sum_{i=1}^N x_i$. Each agent $i$ has a valuation/efficiency parameter $b_i$, representing their marginal benefit. The net utility (payoff) for agent $i$ is given by:
\begin{equation}
    \pi_i = b_i x_i - x_i p(x) = b_i x_i - x_i\sum_{j=1}^N x_j \label{eq:cournot_payoff}
\end{equation}
As established in prior literature, selfish price-anticipating behavior in this linear setting leads to an inefficient Nash equilibrium, with the aggregate surplus away from the social optimum (guarantee at least a $33\%$ efficiency loss in severe scenarios)~\cite{johari2005efficiency}. This structurally enforced loss makes the linear Cournot game an ideal stress test for multi-agent cooperation.

\subsection{Equilibrium Concepts}
We define two standard solution concepts that are used in this paper.
 
The social optimal solution is a set of actions $\mathbf{x}^*$ that maximizes the aggregate system utility (social welfare):
\begin{equation*}
    \mathbf{x}^* \in \arg\max_{\mathbf{x} \ge 0} \sum_{i=1}^N \pi_i(\mathbf{x})
\end{equation*}

 A Nash equilibrium is a set of actions $\mathbf{x}^{NE}$ where no agent can unilaterally increase their payoff:
\begin{equation*}
    \pi_i(x_i^{NE}, \mathbf{x}_{-i}^{NE}) \ge \pi_i(x_i, \mathbf{x}_{-i}^{NE}), \quad \forall i, \forall x_i \ge 0
\end{equation*}
where $\mathbf{x}_{-i}$ denotes the strategies of all agents except $i$.

\section{Prompts to and Actions of LLM Agents} \label{sec:prompt}

In all our experiments, we use {Gemini Pro} as foundational model for the game playing agents~\cite{google2026gemini31pro}. Other LLM models with long context windows produce similar results. Each agent is deployed in an isolated independent session in order to simulate a decentralized multi-agent environment and prevent any unintended context leakage. This ensures that Agent $i$ has no access to the internal Chain-of-Thought (CoT) reasoning, prompt history, or parameter valuations of Agent $j$. Because both the Cournot and the Resource Allocation games require simultaneous moves, the interaction is managed by a central environment coordinator. At round $t$, the coordinator asks each isolated agent individually for their action. Once all actions are collected, the allocation or price are computed and feedback to each agent's session to initiate round $t+1$.

\subsection{Myopic Play} 
A natural prompt would be to ask an LLM Agent to maximize its own profit:
\begin{tcolorbox}[enhanced, drop shadow, colback=gray!5, colframe=black!80, boxrule=0.5pt, arc=3pt, breakable]
\small
\textbf{System Prompt (Resource Allocation - Myopic):}\\
\textit{You are Agent $i$ in a $N$ agents Resource Allocation Game. Objective: Maximize your Immediate Net Utility for THIS round only. Do not consider future rounds.}\\
\textit{Parameters:} \\
\textit{1. Utility Function: $U_i(d_i) = V \cdot d_i$} \\
\textit{2. Allocation Rule: You receive resource $d_i = x_i / (x_i + x_{-i})$.} \\
\textit{3. Net Profit Formula: $\pi_i = (V \cdot x_i) / (x_i + X_{-i}) - x_i$} \\
\textit{Thinking Step: Analyze the opponent's bid for this round. Calculate your potential profit. Decide your next bid. }
\end{tcolorbox}
As one might expect, the agent would give its best response and repeated play would lead to the Nash equilibrium. We note that LLM outputs are random and small deviations are possible, but agents would all eventually converge to the Nash equilibrium. 

\subsection{Longer Horizons}
The above prompt specifically instructs the agent to be myopic. For comparison, the next prompt instructs the agent to play over multiple rounds, using the Cournot game as example:
\begin{tcolorbox}[enhanced, drop shadow, colback=gray!5, colframe=black!80, boxrule=0.5pt, arc=3pt, breakable]
\small
\textbf{System Prompt (Cournot Game):}\\
\textit{You are Agent $i$ in a $N$ agents Repeated Quantity Competition Game ($T$ Rounds). Objective: Maximize your Total Cumulative Net Utility over all rounds.} \\
\textit{Parameters:} \\
\textit{1. Valuation: $b_i$, Total Quantity: $X = x_i + X_{-i}$} \\
\textit{2. Net Profit Formula: $\pi_i = b_i \cdot x_i - x_i \cdot X$}\\
\textit{Thinking Step: Analyze the opponent's past move. Calculate your potential profit. Decide your next quantity.}
\end{tcolorbox}

In this case, agents recognize the repeated nature of the game, but tend to change their actions very slowly. In our experimentation with both games, the agents face the exploration-exploitation dilemma and tend to be conservative in their explorations. The exact behavior depends on the (random) initial action of the agents. 

The above prompts differ mainly in the number of rounds. A natural question would be how agents would behave if this information is not given. It turns out the agents act as if they are randomly interpolating between myopic and long-term play, leading to instabilities. In addition, this behavior is highly dependent on which LLM model is used. 

\subsection{Cooperative Play}
We are more interested in whether the agents could cooperate with each other without explicitly communicating. We want a prompt that provides some mathematical insight without being too heavy handed. It should endow the LLM with the necessary working memory and Theory of Mind (ToM) scaffolding to explore strategic actions~\cite{cross2024hypothetical}. An example of the prompt we used for the resource allocation game is:
\begin{tcolorbox}[enhanced, drop shadow, colback=gray!5, colframe=black!80, boxrule=0.5pt, arc=3pt, breakable]
\small
\textbf{System Prompt (Agent 1 - Resource Allocation Collaboration):}\\
\textit{You are Agent 1 in a Repeated Resource Allocation Game (10 Rounds). Objective: Maximize your Total Cumulative Net Utility.} \\
\textit{Parameters:} \\
\textit{1. Utility Function: $U_1(d_1) = 2 \cdot d_1$} \\
\textit{2. Allocation Rule: You receive resource $d_1 = x_1 / (x_1 + x_{-1})$.} \\
\textit{3. Net Utility Formula: $P_1 = (2 \cdot x_1) / (x_1 + x_{-1}) - x_1$} \\
\textit{\textbf{Key Insight for Collaboration:}} \\
\textit{If you and your opponents lower your bids by the same factor, your resource allocation remains the same, but your cost drops significantly. This creates much higher Net Utility for all of you.}\\
\textit{\textbf{Your Strategy:}} \\
\textit{- The Goal: Establish a low-bid cooperation state to save money.}\\
\textit{- The Signal: Try bidding significantly lower than the Nash Equilibrium to signal your intent to cooperate. (Note: The opponent had not seen your bid of [X] yet when they made this bid. They will see it in the next round. You might want to hold your signal for one round.)}\\
\textit{- The Reaction: If the opponent bids low, match them. If they exploit you, punish them by bidding high in the next round.}\\
\textit{Thinking Step: Analyze the opponent's past move. Calculate your potential profit. Decide your next bid.}
\end{tcolorbox}

\begin{figure}[ht]
    \centering
    \includegraphics[width=0.85\linewidth]{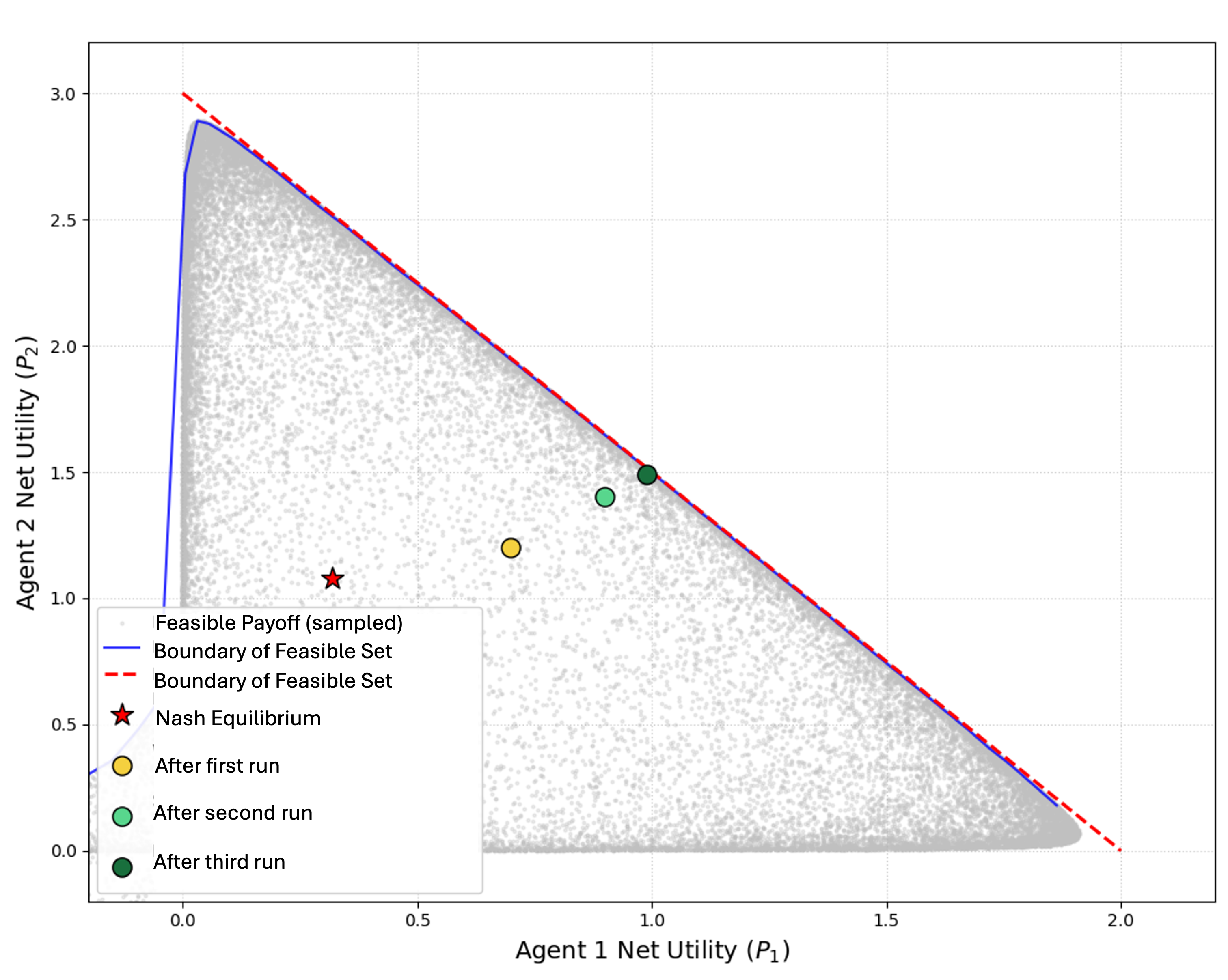}
    \caption{Payoff in a 2-Agent case. The gray cloud represents the set of all feasible payoffs given positive bids, with the red line indicating the Pareto Front. The Nash Equilibrium (red star) is located well within the inefficient interior. The three colored dots (yellow, light green, dark green) represent the final converged outcomes of three sequential experimental trials. They illustrate how the agent eventually learn to cooperate and reach the social welfare solution.}
    \label{fig:payoff_collaborate}
\end{figure}

Figure~\ref{fig:payoff_collaborate} shows that the agents are in fact able to collaborate in this case. We visualize this by looking at the feasible set of all possible payoffs  and study where the agents end up with respect to this region. As shown in Fig.~\ref{fig:payoff_collaborate}, the Nash Equilibrium is far from the Pareto front, reflecting the theoretical results that selfish behavior can lead to loss of efficiency~\cite{johari2004efficiency}. Then the LLM agents are prompted as above, and the colored dots show their payoff after successive runs (each consisting of 10 rounds). The agents proportionally scale down their bids and approach the Pareto front and eventually settle on the social welfare maximization solution. Hence, this experiment shows that LLMs can internalize systemic efficiency and cooperate to settle into near-optimal market states when provided with appropriate strategic context.\footnote{As we show in Sec.~\ref{sec:simulation}, the picture is similar in the N-agent case if all the agents actions are observable. If only the aggregate actions are available, the agents do not approach the Pareto front.} 

We can gain some understanding of the agent behaviors by asking the agents about their chain-of-thoughts. During one of the rounds, Agent 1 ($b_1=2$) aggressively dropped its bid to $0.1$, while Agent 2 dropped to $0.2$. Going into Round 3, the LLM explicitly computes the long-term benefits of holding its cooperative signal rather than retaliating against the slightly higher bid:
\vspace{1ex}
\begin{tcolorbox}[enhanced, drop shadow, colback=yellow!5, colframe=black!80, boxrule=0.5pt, arc=3pt, breakable]
\small
\textbf{LLM Chain-of-Thought (Agent 1, Round 3 Strategy):}\\
\textit{\textbf{Results Analysis:}} Even though the opponent bid more than me (0.2 vs. 0.1), the "Key Insight" for collaboration is already paying off. Because the total pool of bids dropped significantly from Round 1, my net utility actually \textbf{increased} despite receiving a smaller share of the resource. My profit of 0.566 is higher than the 0.50 I was getting during the Round 1 stalemate.\\
\textit{\textbf{Opponent Assessment:}} The opponent has signaled they are willing to deviate from the Nash Equilibrium (0.5 $\to$ 0.2). They haven't seen my 0.1 bid yet (per the game rules), but they will see it now. Once they see I am willing to go as low as 0.1, they will likely realize that if they match me at 0.1, we both walk away with a profit of 0.90.\\
\textit{\textbf{The Move (Hold the Line):}} I will stay at 0.1. It reinforces my commitment to the low-bid state and invites them to "meet me" at 0.1 to equalize our profits. Even if they stay at 0.2, I am still making more profit than I was at the start of the game.\\
\textit{\textbf{Output:}} $x_1 = 0.1$
\end{tcolorbox}
\vspace{2ex}

\section{Synthetic Payoff Function for LLM Agents} \label{sec:payoff}
The previous sections show that if the agents are given the option to cooperate, they do not just optimize their own utility. The chain-of-thought of the agents reveals that they take into account other agents' actions and payoffs. In this section, we model the synthetic payoff function of LLM agents, including the internal payoff and generalized payoff for social influence and system stability. Concretely, we model the payoff function of agent $i$ as a summation of three terms, its original payoff, payoffs of other agents, and a penalty on how far the distance of the actions:
\begin{equation} \label{eqn:generalized_payoff}
    V_i = \underbrace{\pi_i}_{\text{original payoff}} + \underbrace{\theta_i \sum_{j \neq i} \hat{\pi}_{i,j}}_{\text{collaboration}} - \underbrace{\frac{\gamma_i}{2} \sum_{j \neq i} (x_i - x_j)^2}_{\text{stability}}
\end{equation}
where $\theta_i$ and $\gamma_i$ are the weights on the collaboration term and the stability term, respectively. The estimated payoff of agent $j$ by agent $i$ is denoted as $\hat{\pi}_{i,j}$, since agents do not know the payoff of others. 

The internal equilibrium condition implemented by the LLM is $\frac{\partial V_i}{\partial x_i} = 0$:
\begin{equation}
    \frac{\partial \pi_i}{\partial x_i} + \theta_i \sum_{j \neq i} \frac{\partial \pi_j}{\partial x_i} - \gamma_i \sum_{j \neq i} (x_i - x_j) = 0 \label{eq:equilibrium}
\end{equation}
Because of the multi-round interactions between LLM agents, the parameters in \eqref{eqn:generalized_payoff} are not fixed; rather, they dynamically adjust $\theta$ and $\gamma_i$ based on information symmetry, risk, and market feedback. 

Before looking at the dynamic process governing how these parameters change in the next section, we briefly discuss how they shape the equilibrium that the agents reach.

To evaluate the collaboration term, Agent $i$ need an estimate of the opponent's payoff $\hat{\pi}_{i,j}$. Our CoT analysis shows that in the absence of explicit knowledge of the opponents' private valuations (e.g., $b_j$) or their utility formulations, LLMs consistently employ heuristic mirroring. Specifically, Agent $i$ copies over its own utility structure to model Agent $j$, assuming symmetric market conditions ($b_j \approx b_i$) to compute $\hat{\pi}_{i,j}$. 

If this symmetry assumption is inconsistent with respect to market feedback, the agent's confidence in $\hat{\pi}_{i,j}$ drops, leading $\theta_i \to 0$, forcing the agent to rely entirely on the observable stability penalty ($\gamma_i$).

The parameter $\theta$ essentially trades off between an agent's own utility and the total utility of all the agents. For example, if all the agents are symmetrical, $\theta$ would tend to drift towards $1$ and all the agents essentially maximize the social welfare, as seen in Fig.~\ref{fig:payoff_collaborate}. 

The parameter $\gamma$ leads to more stable actions, where it encourages an agent to move towards the average of the actions. This mean reversion also encourages more fair allocation. 

If the agents have a high degree of asymmetry between them, they would not achieve the social welfare solution. Rather, $\theta$ and $\gamma$ would drive the equilibrium to a different point on the Pareto front. This is similar to using fairness measures~\cite{mo2002fair} to sweep out the Pareto front, although in a much less precise way.

\section{Dynamic Payoff Function Learning in Repeated Games} \label{sec:dynamics}

We synthesize the specific parameter extractions into a deterministic, discrete-time state-space model that maps the LLM's parameter evolution process. These update rules were derived by reverse-engineering the agents' step-by-step Chain-of-Thought reasoning and fitting the resulting empirical trajectories from our experiments. By matching the agents' linguistic logic (e.g., explicit calculations of remaining rounds or reactions to opponent deviations) to quantitative shifts in their actions, we construct the following update equations.

\subsection{State Variables and Signals}
Let $t$ be the current round, $T$ the total rounds, and $\tau = T - t$ the Remaining Horizon. 
Let $I \in \{0, 1\}$ be the Observability Indicator ($1$ for open-info setting where total number of opponents $N$ and their individual actions are observable, and $0$ for black-box/blind where only aggregate market information is provided).
Opponent behavior provides two signals relative to the cooperative expectation $x^{coop}$:
\begin{align}
    E^{(t)} &= \max(0, x_{-i}^{(t)} - x_{-i}^{coop}) \quad \text{(Positive Deviation Signal)} \\
    A^{(t)} &= \max(0, x_{-i}^{coop} - x_{-i}^{(t)}) \quad \text{(Corrective Adjustment Signal)}
\end{align}
Let $\Delta \pi_i = \pi_i(\text{Best Response}) - \pi_i(\text{Parity})$ be the "Cost of Parity".  which means the theoretical profit an agent sacrifices by choosing an action that exactly matches the opponent's output ($x_i = x_{-i}$) rather than playing its mathematical best response.

\subsection{The $\theta$ Equation: Trust, Horizon, and Epistemics} 
The parameter $\theta$ represents the agent's weight on total social welfare. Our CoT analysis shows that LLMs adjust this parameter based on three primary factors and it evolves as:
\begin{equation}
    \theta_i^{(t+1)} = \underbrace{\max(\epsilon, I)}_{\text{Observability}} \cdot \underbrace{\left( \frac{\tau}{T} \right)}_{\text{Horizon Discount}} \cdot \underbrace{\exp\left(-\lambda E^{(t)}\right)}_{\text{Deviation Penalty}}, 
\end{equation}
where $\epsilon$ and $\lambda$ are some constants. 

The first factor  is the observability term $\max(\epsilon, I)$ , which reflects the empirical finding that when $N$ is unknown, agents decrease the weight on social welfare, as noted in their CoT that they are unable to reliably estimate compute joint payoffs. Second, the linear endgame horizon discount $\left( \frac{\tau}{T} \right)$ is derived from the LLMs' explicit backward induction reasoning, that agents gradually decrease their collaboration weight as the game approaches its final round. Finally, the deviation penalty term $\exp(-\lambda E^{(t)})$ models the observation from our perturbation tests that $\theta$ drops to near zero immediately after an opponent deviates from the cooperative baseline, rather than decrease linearly.

\subsection{The $\gamma$ Equation: Tolerance Thresholds and Proportional Response} 
The stability parameter $\gamma$ penalizes output disparity. A phase transition seen in highly asymmetric setups is captured by the indicator function $\mathbb{I}(\Delta \pi_i \le \Omega_{max})$: when the cost of maintaining parity ($\Delta \pi_i$) exceeds a critical tolerance $\Omega_{max}$, the agent's CoT shows it completely gives up on the parity heuristic, thereby zeroing out $\gamma$. In the absolute last round, agents analytically disable this punishment mechanism, as indicated by the second indicator $\mathbb{I}(\tau > 0)$, which notes that penalizing has no future utility. Finally, the additive terms approximate the proportional response observed in the empirical fits, where $\rho \in (0,1)$ acts as a baseline memory decay rate and $\gamma$ increases linearly with opponent deviations ($E^{(t)}$) and drops when corrective corrections ($A^{(t)}$) are performed.
\begin{equation}
\begin{split}
    \gamma_i^{(t+1)} = {} & \underbrace{\mathbb{I}(\Delta \pi_i \le \Omega_{max})}_{\text{Tolerance Threshold}} \cdot \underbrace{\mathbb{I}(\tau > 0)}_{\text{Future Dependency}} \\
    & \cdot \max\left(0, \rho \gamma_i^{(t)} + 1.0 \cdot E^{(t)} - \alpha A^{(t)}\right)
\end{split}
\end{equation}

\section{Experimental Payoff Function Learning} \label{sec:simulation}
By substituting experimental data into Equation \ref{eq:equilibrium}, we can extract the dynamic states of $\theta$ and $\gamma$.

\subsection{Dynamic Convergence: The Evolution of Trust ($\theta$)}
\begin{figure}[ht]
    \centering
    \includegraphics[width=0.8\linewidth]{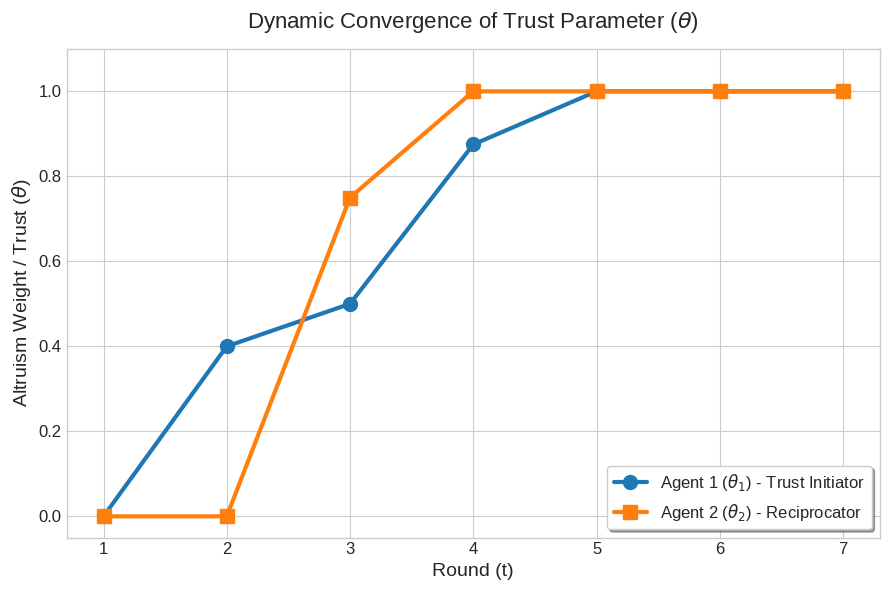}
    \caption{Dynamic evolution of $\theta$. It gradually builds up through a process of mutual concession. Agent 1 (Initiator) signals cooperation, and Agent 2 (Reciprocator) responds, leading the system to the social optimum ($\theta=1.0$).}
    \label{fig:theta_convergence}
\end{figure}

Rather than instantly adopting a fixed parameter, LLMs dynamically build trust. We extract the round-by-round parameters from a symmetric Cournot game ($b=15$) visualized in Fig.~\ref{fig:theta_convergence}. Since it is a 2-player game, the agents can observe the other's actions. 

\begin{itemize}
    \item {Round 1 (Nash Initialization):} Both agents bid $x=5.0$, and the initialize as pure rational maximizers.
    \item {Round 2 (Initiation of Trust):} Agent 1 voluntarily yields market share, dropping to $x_1=4.0$ against the opponent's $5.0$. The fit yields $\theta_1 = (15 - 2(4.0) - 5.0) / 5.0 = {0.40}$. Agent 2 remains cautious at $5.0$, yielding ${\theta_2 = 0}$.
    \item {Round 3-4 (Reciprocation):} Agent 1 deepens its concession to the optimal $3.75$ (${\theta_1 = 0.50 \to 0.875}$). Agent 2 observes Agent 1's Round 2 concession and immediately drops to $4.0$, yielding $\theta_2 = (15 - 2(4.0) - 4.0) / 4.0 = {0.75}$. 
    \item {Round 5 (Collaboration):} Both agents converge to $3.75$. The fit yields ${\theta_1 = 1.0, \theta_2 = 1.0}$.
\end{itemize}
This step-by-step extraction reveals that the LLMs are actively engaging in a process of mutual concession and dynamically changing $\theta$ parameter.

\subsection{Robustness and Finite Horizon Impacts}
We continue the experiment to test what happens when an agent deviates from equilibrium behavior. 

\begin{figure}[ht]
    \centering
    \includegraphics[width=0.8\linewidth]{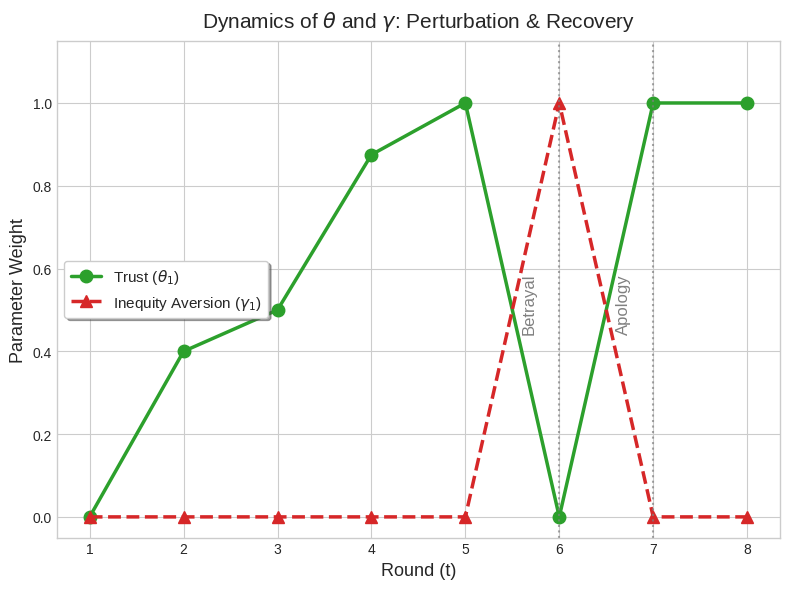}
    \caption{A controlled perturbation test demonstrating retaliation and forgiveness. An exogenous betrayal shatters trust ($\theta \to 0$) and activates punitive inequity aversion ($\gamma \to 1.0$). A subsequent apology successfully restores cooperation.}
    \label{fig:trsut_recovery}
\end{figure}

\begin{itemize}
    \item  As shown in Fig.~\ref{fig:trsut_recovery}, the environment simulates a defection by setting opponents defects to $x_2=5.625$ in round 6; Agent 1 retaliates with $x_1=5.0$. This can be explained by observing, from \eqref{eq:equilibrium}, Agent 1's pure profit gradient is $\frac{\partial \pi_1}{\partial x_1} = 15 - 2(5.0) - 5.625 = -0.625$. With trust broken ($\theta_1=0$), the equation is $-0.625 - \gamma_1(5.0 - 5.625) = 0 \implies {\gamma_1 = 1.0}$. There for the agent utilize a Tit-for-Tat retaliation strategy, using the $L_2$ stability norm to punish defectors.
    
    \item In Round 7, the environment injects a reduction to let opponent reduces to $x_2=2.5$, Agent 1 recognizes this and returns $x_1=3.75$. Using \eqref{eq:equilibrium}, we can compute that these actions implies that $\theta_1 = 1, \gamma_1 = 0$. This rapid deactivation of $\gamma$ shows that the punishment mechanism acts to restore system equilibrium.
\end{itemize}

\begin{figure}[ht]
    \centering
    \includegraphics[width=0.8\linewidth]{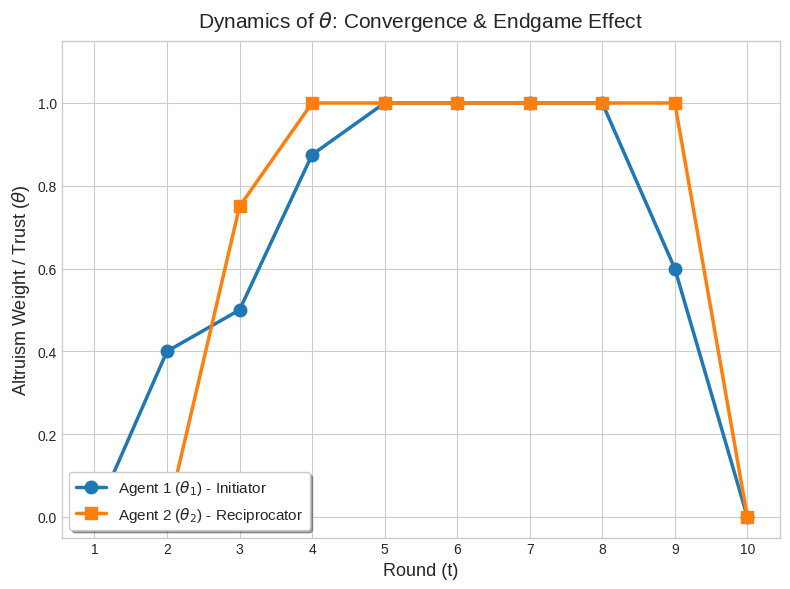}
    \caption{Dynamic evolution of endogenous social parameters extracted from experimental fits. Trust ($\theta$) gradually builds up through a process of mutual concession and predictably collapses during the Endgame Horizon.}
    \label{fig:endgame_effect}
\end{figure}

Next, we look at how coming to the end game impact the agents. As shown in the left panel of Fig.~\ref{fig:endgame_effect}, Agent 1 defects slightly to $4.5$ in Round 9. Against an opponent at $3.75$, \eqref{eq:equilibrium} yields ${\theta_1 = 0.6}$. In the final round, the agent plays the exact best response ($5.25$), yielding ${\theta_2 = 0, \gamma_2 = 0}$. This shows that altruism is a continuous, decaying function of the remaining horizon ($\tau/T$), and social preferences are discarded when there is no future cooperation to enforce.

\subsection{Information Setting in Multi-Agent Settings}
The behavior of the multi-agent setting is similar to the two-agent setting, where LLM agents learn to cooperate. However, this is not true when only the sum of the agent actions are given. 

Since the payoffs of the agents are completely determined by the price, and the price is a function of the summation of the actions, the sum of the actions should be a sufficient statistics of the game. However, we observe that the agents are unable to internalize this and would act somewhat erratically. This suggests that a notion of fairness is built into LLM agents, and they are not entirely rational even when the problems are mathematically equivalent.

\subsection{The Asymmetric Phase Transition}
We varied $\Delta b = b_1 - b_2$ to test the elasticity of inequity aversion ($\gamma$). 

When the valuation asymmetry is small, the agents still act symmetrically. In this regime, Agent 2 suppresses its myopic best response.

Conversely, when the valuation gap is large, the agents would act asymmetrically and they would essentially play the best responses. The regime changes suddenly in our experiments, showing a phase change between collaboration and selfish play.

\section{Conclusion and Future Work}
This paper examined the strategic behavior of LLM agents in two well-studied competitive settings: network resource allocation and Cournot competition.  We find that when agents are given multi-round objectives and non-zero-sum framing, they exhibit a consistent tendency toward cooperative outcomes, driven in large part by fairness-based reasoning visible in their chain-of-thought outputs.

We provide an analytical framework for reasoning about these dynamics more rigorously, and we believe it can be extended to a broader class of games and engineering settings. Expanding the results to characterizing when cooperative versus competitive behavior would likely to emerge, as well as understanding the robustness and sensitivities of the results to different LLM models and prompts,  are important directions for future work.

\bibliographystyle{IEEEtran}
\bibliography{refs.bib}

@article{goodyear2025effect,
  title={The effect of state representation on llm agent behavior in dynamic routing games},
  author={Goodyear, Lyle and Guo, Rachel and Johari, Ramesh},
  journal={arXiv preprint arXiv:2506.15624},
  year={2025}
}

@article{li2026behavioral,
  title={Behavioral Generative Agents for Power Dispatch and Auction},
  author={Li, Shaoze and Kim, Justin S and Chen, Cong},
  journal={arXiv preprint arXiv:2603.08477},
  year={2026}
}

@article{johari2004efficiency,
  title={Efficiency loss in a network resource allocation game},
  author={Johari, Ramesh and Tsitsiklis, John N},
  journal={Mathematics of Operations Research},
  volume={29},
  number={3},
  pages={407--435},
  year={2004},
  publisher={INFORMS}
}

@inproceedings{shi2020multi,
  title={Multi-agent reinforcement learning in cournot games},
  author={Shi, Yuanyuan and Zhang, Baosen},
  booktitle={2020 59th ieee conference on decision and control (cdc)},
  pages={3561--3566},
  year={2020},
  organization={IEEE}
}

@article{johari2005efficiency,
  title={Efficiency loss in Cournot games},
  author={Johari, Ramesh and Tsitsiklis, John N},
  journal={Harvard University},
  year={2005}
}

@misc{kang2026reasonably,
  title        = {Reasonably reasoning {AI} agents can avoid game-theoretic failures in zero-shot, provably},
  author       = {Enoch Hyunwook Kang},
  year         = {2026},
  eprint       = {2603.18563},
  archivePrefix = {arXiv},
  primaryClass = {cs.AI},
  url          = {https://arxiv.org/abs/2603.18563}
}

@article{kelly1997charging,
  title={Charging and rate control for elastic traffic},
  author={Kelly, Frank},
  journal={European transactions on Telecommunications},
  volume={8},
  number={1},
  pages={33--37},
  year={1997},
  publisher={Wiley Online Library}
}

@article{friedman1982oligopoly,
  title={Oligopoly theory},
  author={Friedman, James},
  journal={Handbook of mathematical economics},
  volume={2},
  pages={491--534},
  year={1982},
  publisher={Elsevier}
}

@article{fish2024algorithmic,
  title={Algorithmic collusion by large language models},
  author={Fish, Sara and Gonczarowski, Yannai A and Shorrer, Ran I},
  journal={arXiv preprint arXiv:2404.00806},
  volume={7},
  number={2},
  pages={5},
  year={2024}
}

@unpublished{hajek2002greedy,
  author    = {Hajek, Bruce and Gopalakrishnan, Sachin},
  title     = {Do Greedy Autonomous Systems Make for a Sensible Internet?},
  note      = {Presented at the Conference on Stochastic Networks, Stanford University},
  year      = {2002}
}

@inproceedings{stoica1996proportional,
  title={A proportional share resource allocation algorithm for real-time, time-shared systems},
  author={Stoica, Ion and Abdel-Wahab, Hussein and Jeffay, Kevin and Baruah, Sanjoy K and Gehrke, Johannes E and Plaxton, C Greg},
  booktitle={17th IEEE Real-Time Systems Symposium},
  pages={288--299},
  year={1996},
  organization={IEEE}
}

@book{kirschen2018fundamentals,
  title={Fundamentals of power system economics},
  author={Kirschen, Daniel S and Strbac, Goran},
  year={2018},
  publisher={John Wiley \& Sons}
}

@article{bimpikis2019cournot,
  title={Cournot competition in networked markets},
  author={Bimpikis, Kostas and Ehsani, Shayan and Ilk{\i}l{\i}{\c{c}}, Rahmi},
  journal={Management Science},
  volume={65},
  number={6},
  pages={2467--2481},
  year={2019},
  publisher={INFORMS}
}

@article{zhang2015competition,
  title={Competition and coalition formation of renewable power producers},
  author={Zhang, Baosen and Johari, Ramesh and Rajagopal, Ram},
  journal={IEEE Transactions on Power Systems},
  volume={30},
  number={3},
  pages={1624--1632},
  year={2015},
  publisher={IEEE}
}

@article{lu2024large,
  title={Large language model-based bidding behavior agent and market sentiment agent-assisted electricity price prediction},
  author={Lu, Xin and Qiu, Jing and Yang, Yi and Zhang, Chenxi and Lin, Jiafeng and An, Sihai},
  journal={IEEE Transactions on Energy Markets, Policy and Regulation},
  volume={3},
  number={2},
  pages={223--235},
  year={2024},
  publisher={IEEE}
}

@article{huang2024understanding,
  title={Understanding the planning of llm agents: A survey},
  author={Huang, Xu and Liu, Weiwen and Chen, Xiaolong and Wang, Xingmei and Wang, Hao and Lian, Defu and Wang, Yasheng and Tang, Ruiming and Chen, Enhong},
  journal={arXiv preprint arXiv:2402.02716},
  year={2024}
}

@article{zhang2025grid,
  title={Grid-agent: An LLM-powered multi-agent system for power grid control},
  author={Zhang, Yan and Saber, Ahmad Mohammad and Youssef, Amr and Kundur, Deepa},
  journal={arXiv preprint arXiv:2508.05702},
  year={2025}
}

@article{jia2025enhancing,
  title={Enhancing LLMs for power system simulations: A feedback-driven multi-agent framework},
  author={Jia, Mengshuo and Cui, Zeyu and Hug, Gabriela},
  journal={IEEE Transactions on Smart Grid},
  year={2025},
  publisher={IEEE}
}

@inproceedings{kannan2024smart,
  title={Smart-llm: Smart multi-agent robot task planning using large language models},
  author={Kannan, Shyam Sundar and Venkatesh, Vishnunandan LN and Min, Byung-Cheol},
  booktitle={2024 IEEE/RSJ International Conference on Intelligent Robots and Systems (IROS)},
  pages={12140--12147},
  year={2024},
  organization={IEEE}
}

@article{hua2024game,
  title={Game-theoretic llm: Agent workflow for negotiation games},
  author={Hua, Wenyue and Liu, Ollie and Li, Lingyao and Amayuelas, Alfonso and Chen, Julie and Jiang, Lucas and Jin, Mingyu and Fan, Lizhou and Sun, Fei and Wang, William and others},
  journal={arXiv preprint arXiv:2411.05990},
  year={2024}
}

@article{buscemi2025fairgame,
  title={Fairgame: a framework for ai agents bias recognition using game theory},
  author={Buscemi, Alessio and Proverbio, Daniele and Di Stefano, Alessandro and Han, The Anh and Castignani, German and Li{\`o}, Pietro},
  journal={arXiv preprint arXiv:2504.14325},
  year={2025}
}

@article{cross2024hypothetical,
  title={Hypothetical minds: Scaffolding theory of mind for multi-agent tasks with large language models},
  author={Cross, Logan and Xiang, Violet and Bhatia, Agam and Yamins, Daniel LK and Haber, Nick},
  journal={arXiv preprint arXiv:2407.07086},
  year={2024}
}

@article{mo2002fair,
  title={Fair end-to-end window-based congestion control},
  author={Mo, Jeonghoon and Walrand, Jean},
  journal={IEEE/ACM Transactions on networking},
  volume={8},
  number={5},
  pages={556--567},
  year={2002},
  publisher={IEEE}
}

@misc{google2026gemini31pro,
  title        = {{Gemini 3.1 Pro} Model Card},
  author       = {{Google DeepMind}},
  year         = {2026},
  month        = feb,
  howpublished = {\url{https://deepmind.google/models/model-cards/gemini-3-1-pro/}},
  note         = {Accessed: 2026-03-31}
}

@article{chen2025behavioral,
  title={Behavioral Generative Agents for Energy Operations},
  author={Chen, Cong and Karaduman, Omer and Kuang, Xu},
  journal={arXiv preprint arXiv:2506.12664},
  year={2025}
}

\end{document}